\documentclass[aps,twocolumn,floatfix,showpacs,amsmath,amsfonts]{revtex4}

\usepackage{graphicx}
\usepackage{bm}

\begin{document}

\title{Dynamics and stability of Bose-Einstein condensates with 
attractive $1/r$ interaction}

\author{Holger Cartarius}
\email{Holger.Cartarius@itp1.uni-stuttgart.de}
\author{Toma\v z Fab\v ci\v c}
\author{J\"org Main}
\author{G\"unter Wunner}
\affiliation{Institut f\"ur Theoretische Physik 1, Universit\"at Stuttgart,
  70550 Stuttgart, Germany}
\date{\today}

\begin{abstract}
The time-dependent extended Gross-Pitaevskii equation for Bose-Einstein
condensates with attractive $1/r$ interaction is investigated with both
a variational approach and numerically exact calculations.
We show that these condensates exhibit signatures known
from the nonlinear dynamics of autonomous Hamiltonian systems.
The two stationary solutions created in a tangent bifurcation at a
critical value of the scattering length are identified as elliptical
and hyperbolical fixed points, corresponding to stable and
unstable stationary states of the condensate. 
The stable stationary state is surrounded by elliptical islands, corresponding
to condensates periodically
oscillating in time, whereas condensate wave functions in the unstable region
undergo a collapse within finite time.
For negative scattering lengths below the tangent bifurcation no stationary
solutions exist, i.e., the condensate is always unstable and collapses.
\end{abstract}

% \keywords{}

\maketitle

\section{Introduction}
\label{sec:intro}
Bose condensates with atomic interactions, in addition to the
contact interaction, open the 
possibility to study the properties of degenerate quantum gases in which the
relative strengths of long- and short-range interactions can be continuously
adjusted by tuning the contact interaction via a Feshbach resonance.
In particular, the Bose-Einstein condensation of neutral atoms with
electromagnetically induced attractive $1/r$ interaction has been proposed.
According to O'Dell et al.\ \cite{ODe00} ``monopolar'' quantum gases could be
realized by a combination of 6 appropriately arranged ``triads'' of intense
off-resonant laser beams. In that arrangement, the  $1/r^3$ interactions of
the retarded dipole-dipole interaction of neutral atoms in the presence of
intense electromagnetic radiation are averaged out in the near-zone limit,
while the weaker $1/r$ interaction is retained. An outstanding feature of
this type of long-range interaction is that for attractive contact
interaction stable Bose-Einstein condensates are predicted that are
\emph{self-bound} (without an additional trap). Despite the fact
that a self-bound condensate with attractive $1/r$ interaction has not
been realized in an experiment so far, the physical parameters necessary
for creating it experimentally were discussed in detail by Giovanazzi et
al.\ \cite{Gio01a}, who also obtained a theoretical prediction for the number
of atoms in the self-bound condensate. Furthermore, the existence of stable
monopole and quadrupole oscillations around the stationary ground state have
been demonstrated and analyzed with analytical and numerical calculations
\cite{Gio01b}.

The realization of self-bound condensates with attractive
$1/r$ interaction in experiments remains a challenging task.
However, experimental realization has been achieved for a similar
system with a long-range interaction, viz.\ dipolar Bose-Einstein
condensates. Here, the long-range dipole-dipole interaction appears
next to the short-range contact term. The system has attracted much
attention in recent years 
\cite{santos00,baranov02,goral02a,goral02b,giovanazzi03}
and the achievement of Bose-Einstein condensation
in a gas of chromium atoms \cite{griesmaier05}, with a large dipole moment,
has opened the way to promising experiments on dipolar quantum gases
\cite{stuhler05}. For example, the experimental observation of the dipolar
collapse of a quantum gas has recently been reported by Koch et al.\
\cite{koch08}, which sets in when the contact interaction is reduced below
some critical value.

Bose-Einstein condensates with attractive $1/r$ interaction exhibit a
similar collapse scenario. Collapse of the self-bound condensates again only
sets in below some critical value of the scattering length. 
It was recently shown \cite{Pap07} that these critical values in fact
correspond to bifurcation points where two solutions of the Gross-Pitaevskii
equation disappear: one the true ground state, the other a
collectively excited state. It was also demonstrated \cite{Car07b} that at
the bifurcation the two stationary solutions exhibit the typical structure
known from studies of \emph{exceptional points}
\cite{Kato66,Hei99,Hei04,Gue07,Car07a} in open quantum systems described by
non-Hermitean Hamiltonians. At the exceptional points both the energies
\emph{and} the corresponding wave functions are identical, a situation which
is forbidden for bound states of the linear Schr\"odinger equation (which
have to be orthogonal) but is possible here because of the nonlinearity of
the Gross-Pitaevskii equation.

Using a complex continuation of the Gross-Pitaevskii equation the existence
of complex eigenstates at (real) negative scattering lengths below the
bifurcation point has also been revealed \cite{Car07b}.
The physical interpretation of these states is the collapse (or explosion)
of the condensate with a decay rate given by the imaginary part of the
complex eigenvalues of the chemical potential.

It is the purpose of this paper to analyze self-bound
spherically symmetric  Bose condensates with attractive $1/r$ interaction
in the vicinity of the bifurcation points 
 from the point of view of nonlinear dynamics, and to investigate the time
evolution of arbitrary condensate wave functions. We do this by solving the
time-dependent Gross-Pitaevskii equation  both by means of a variational
approach with time-dependent Gaussian wave packets and by exact numerical
computations using the split-operator method. We will show that of the two 
stationary solutions created in a tangent bifurcation one is dynamically
stable and the other unstable, corresponding to elliptic and hyperbolic fixed
points, respectively. The stable state is surrounded by solutions periodically
oscillating in time, whereas wave functions in the unstable region undergo
a collapse within finite time.
Below the tangent bifurcation no stationary solutions exist, i.e., the
condensate is always unstable and collapsing. 

The special appeal of investigations of the properties of spherically symmetric
self-trapped Bose condensates with attractive $1/r$ interaction lies
in the fact that the extremization of the variational mean-field energy 
can be carried out fully analytically.
The reason is that in absence of a trap the extremization conditions 
for the mean-field energy become quadratic equations, while with
a trap potential the equations become at least polynomials of order 5, caused
by the combination of the trap and contact interaction terms, regardless of
whether or not a long-range interaction is present, and of its type. Therefore
generic properties of Bose condensates can be elucidated in a very simple and
transparent way, and later be checked in numerical calculations and in more
complex systems such as dipolar Bose condensates.

The evolution of a wave function is determined by the extended
Gross-Pitaevskii equation for self-trapped Bose-Einstein condensates with
attractive $1/r$ interaction without external trap potential, which reads
\begin{multline}
 {i}\frac{d}{{d}t} \psi(\bm r,t) = \bigg ( -\Delta 
   + 8\pi N a \left | \psi(\bm{r},t)\right |^2 \\
   - 2 N \int d^3 \bm{r}' \frac{\left | \psi(\bm{r}',t)\right |^2}
   {\left | \bm{r} - \bm{r}' \right |} \bigg ) \psi(\bm{r},t) \; ,
\label{eq:extended_GP}
\end{multline}
where the natural ``atomic'' units introduced in Ref.\ \cite{Pap07} were used.
Lengths are measured in units of a ``Bohr radius'' $a_u=\hbar^2/(m u)$,
energies in units of a ``Rydberg energy'' $E_u=\hbar^2/(2 m a_u^2)$,
and times in units of $t_u = \hbar/E_u$, where $u$ determines the strength of
the atom-atom interaction \cite{ODe00}, and $m$ is the mass of one boson.
The number of bosons is $N$.

The paper is organized as follows.
In Sec.\ \ref{sec:TDVP} the basic equations resulting from a time-dependent
variational principle applied to the Gross-Pitaevskii equation are
presented, where the solution of the nonlinear partial differential equation
is reduced to a set of ordinary first order differential equations. 
The linear stability of the fixed points is investigated with variational as
well as numerically exact calculations in Sec.\ \ref{sec:linstab}.
In Sec.\ \ref{sec:dynamics} the dynamics of the condensate obtained from
variational and from numerically exact computations is investigated and the
results are compared. Conclusions are drawn in Sec.\ \ref{sec:disc}.

\section{Time-dependent variational principle}\label{sec:TDVP}
We apply the time-dependent variational principle (TDVP)
\cite{Dir30,McL64} 
to solving the time-dependent Gross-Pitaevskii (GP) equation. Exploiting
the scaling properties presented in Ref.\ \cite{Pap07} and introducing the
scaled variables
\begin{equation}
  (\tilde{\bm{r}}, \tilde{a}, \tilde{t}, \tilde{\psi})
  = (Nr, N^2a, N^2 t, N^{-3/2}\psi)\; ,
  \label{eq:scaling}
\end{equation}
we bring the system in the form
\begin{equation}
  i \frac{d}{d\tilde{t}} \tilde{\psi}(\tilde{\bm{r}},\tilde{t}) = 
  \tilde{H} \tilde{\psi}(\tilde{\bm{r}},\tilde{t})
  = \left[-\Delta_{\tilde{\bm{r}}} + \tilde{V}_c + \tilde{V}_u \right] 
  \tilde{\psi}(\tilde{\bm{r}},\tilde{t}) \; ,
  \label{eq:scaled_extended_GP}
\end{equation}
where the potentials
\begin{subequations}
  \begin{align}
    \tilde{V}_c &= 8\pi \tilde{a}|\tilde{\psi}(\tilde{\bm{r}},\tilde{t})|^2 
    \; ,\\
    \tilde{V}_u &= - 2 \int d^3 \tilde{\bm{r}}' 
    \frac{|\tilde{\psi}(\tilde{\bm{r}}',\tilde{t})|^2}
    {| \tilde{\bm{r}} - \tilde{\bm{r}}' |} 
    \label{eq:integral_Vu}
  \end{align}
\end{subequations}
depend on the coordinates and the wave function, i.e., $\tilde{H}$ is a
\emph{nonlinear} operator. The scaling reveals that the system has only
one external parameter, viz., $\tilde{a} = N^2 a/a_u$ \cite{Pap07}. 
If not stated
otherwise, we use the scaled variables throughout the rest of the paper and
omit the tilde in what follows.

An approximate solution $\psi(t)$ on a given manifold in Hilbert space 
is obtained by minimizing the quantity 
$I = ||i \phi(t) -H \psi(t)||^2 $
with respect to $\phi$ only, and then setting $\dot\psi\equiv\phi$.
This means that the time-dependent variational principle does not depend
on properties of the operator $H$, i.e., the method can be applied to 
linear and nonlinear operators in the same way. Let the
wave function be parametrized by a set of complex time-dependent functions
${\bm z}(t)=\{z_1(t),\dots,z_{n}(t)\}$, i.e., $\psi(t)=\psi({\bm z}(t))$. 
The TDVP yields a set of ordinary differential equations for the motion
of ${\bm z}(t)$
\begin{equation}
 K \dot{{\bm z}}= -i {\bm{h}} 
\label{eq:gg}
\end{equation}
with the matrix $K$ and the vector $\bm{h}$ of the $n$-dimensional system
\eqref{eq:gg} given by
\begin{equation}
 K = \left\langle \frac{\partial \psi}{ \partial{{\bm z}}}
   \Big|\frac{\partial\psi}{\partial{\bm z}} \right\rangle \; , \quad
 {\bm{h}} = \left\langle \frac{\partial \psi} { \partial{{\bm z}}}
   \Big|H \Big|\psi \right\rangle \; .
\label{eq:Kh}
\end{equation}
In time-independent calculations for condensates with $1/r$ interaction
Gaussian wave functions have been used as an ansatz for the two solutions
created in the tangent bifurcation \cite{ODe00,Pap07}.
To apply the TDVP to the Gross-Pitaevskii equation \eqref{eq:extended_GP}
we choose as a test function a radially symmetric Gaussian wave packet
\begin{equation}
 \psi(r,t) = e^{i(A r^2+\gamma)} = e^{-(A_i r^2+\gamma_i)+i(A_r r^2+\gamma_r)}
\label{eq:psi}
\end{equation}
with the time-dependent variational parameters
$\bm z(t)=\{A(t),\gamma(t)\}=\{A_r(t)+iA_i(t),\gamma_r(t)+i\gamma_i(t)\}$.
A similar time-dependent Gaussian trial function has  
been applied in studies of  the 
dynamics of the Gross-Pitaevskii equation without $1/r$ interaction 
\cite{Per97}.
Evaluation of $K$ and $\bm{h}$ in \eqref{eq:gg} and \eqref{eq:Kh} yields
the ordinary differential equations
\begin{subequations}
  \begin{align}
    \label{eq:dgl_A_1}
    \dot A &= -4 A^2 - \frac{1}{2}V_2 \; , \\
    \label{eq:dgl_gamma_1}
    \dot\gamma &= 6iA - v_0 \; ,
  \end{align}
\end{subequations}
where $v_0$ and $V_2$ are given as solutions of the two linear equations
\begin{subequations}
  \begin{align}
    \langle\psi|\psi\rangle v_0 + \frac{1}{2}\langle\psi|r^2|\psi\rangle V_2
    &= \langle\psi|V_c+V_u|\psi\rangle \; , \\
    \langle\psi|r^2|\psi\rangle v_0 + \frac{1}{2}\langle\psi|r^4|\psi\rangle V_2
    &= \langle\psi|r^2(V_c+V_u)|\psi\rangle \; ,
  \end{align}
\end{subequations}
with
\begin{subequations}
  \begin{align}
    \langle\psi|\psi\rangle
    &= \frac{\pi^{3/2}}{2\sqrt{2}} e^{-2\gamma_i}A_i^{-3/2} \; ,\\
    \langle\psi|r^2|\psi\rangle
    &= \frac{3\pi^{3/2}}{8\sqrt{2}} e^{-2\gamma_i}A_i^{-5/2} \; ,\\
    \langle\psi|r^4|\psi\rangle
    &= \frac{15\pi^{3/2}}{32\sqrt{2}} e^{-2\gamma_i}A_i^{-7/2} \; ,\\
    \langle\psi|V_c|\psi\rangle
    &= \pi^{5/2} a e^{-4\gamma_i}A_i^{-3/2} \; ,\\
    \langle\psi|V_u|\psi\rangle
    &= -\frac{\pi^{5/2}}{2} e^{-4\gamma_i}A_i^{-5/2} \; ,\\
    \langle\psi|r^2 V_c|\psi\rangle
    &= \frac{3\pi^{5/2}}{8} a e^{-4\gamma_i}A_i^{-5/2} \; ,\\
    \langle\psi|r^2 V_u|\psi\rangle
    &= -\frac{5\pi^{5/2}}{16} e^{-4\gamma_i}A_i^{-7/2} \; .
  \end{align}
\end{subequations}
Inserting $v_0$ and $V_2$ in \eqref{eq:dgl_A_1} and \eqref{eq:dgl_gamma_1}
leads to the differential equations
\begin{subequations}
  \begin{align}\label{eq-adot}
    \dot A &= -4A^2 + 2\sqrt{2}\pi e^{-2\gamma_i}
    \left(a A_i-\frac{1}{6}\right) \; , \\
\label{eq-gammadot}    
\dot\gamma &= 6iA + \frac{\pi e^{-2\gamma_i}}{2\sqrt{2}\,A_i}
    \left(5-14 a A_i\right) \; .
  \end{align}
\end{subequations}
The imaginary part of Eq. \eqref{eq-gammadot} can be integrated 
analytically,
\begin{equation}
  \gamma_i(t)=-\frac{3}{4}\ln\frac{2A_i(t)}{\pi} \; ,
\end{equation}
and guarantees the normalization condition $||\psi||^2=1$.

The final form of the equations of motion in real number representation is
given by
\begin{subequations}
  \label{eq:Adot_1}
  \begin{align}
    \label{eq:Ardot_1}
    \dot A_r &= -4(A_r^2-A_i^2)
    + \frac{8}{\sqrt{\pi}}A_i^{3/2} \left(a A_i-\frac{1}{6}\right) \;,\\
    \label{eq:Aidot_1}
    \dot A_i &= -8 A_r A_i \; , \\
    \label{eq:grdot_1}
    \dot\gamma_r &= -6A_i + \frac{1}{\sqrt{\pi}}\sqrt{A_i}
    \left(5-14 a A_i\right) \; ,
  \end{align}
\end{subequations}
which are solved under the initial conditions
$A_r(0)=\gamma_r(0)=0$, $A_i(0)>0$
for an initially real valued Gaussian wave packet.

\section{Linear stability of the bifurcating states} \label{sec:linstab}

In this section we study the stability of the two stationary eigenstates 
that are born in the tangent bifurcation. We do this first for the variational
solutions, and then demonstrate how the essential features
carry over to the analysis of the numerically exact solutions.

\subsection{Stability of the variational solutions}

\begin{figure}[tb]
  \includegraphics[width=\columnwidth]{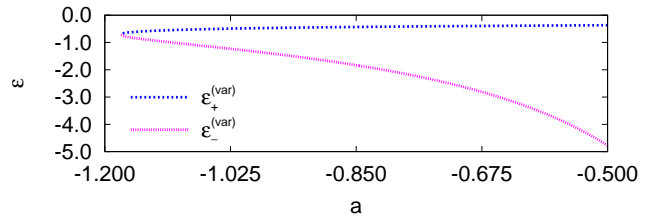}
  \caption{\label{fig:stationary}(Color online)
    Chemical potentials of the ground state ($\varepsilon_+$) and the
    nodeless excited state ($\varepsilon_-$) in the variational calculation.
    They emerge in a tangent bifurcation at a critical value of the scaled
    scattering length $a_\mathrm{cr}=-3\pi/8=-1.1780$ \cite{Pap07}.}
\end{figure}

In the time-dependent variational approach, the two stationary states of the
Gross-Pitaevskii equation \eqref{eq:scaled_extended_GP} appear as the
time-independent solutions (fixed points) of the equations of motion
\eqref{eq:Adot_1}. Requiring $\dot{A}_i = 0$  and $\dot{A}_r = 0 $ immediately
leads to
\begin{subequations}
  \begin{align}
    \hat{A}_r &= 0\; , 
    \label{eq:fixed_points_Ar} \\
    \hat{A}_i &= \frac{1}{6 a} + \frac{\pi}{8 a^2} \left(1\pm\sqrt{1+\frac{8 a}
        {3\pi}} \right) \; .
    \label{eq:fixed_points_Ai}
  \end{align}
\end{subequations}
The vanishing of the real part of $\hat{A}$ implies that the state indeed
is a stationary Gaussian. The scaled chemical
potentials $\varepsilon$ are given
by the negative time derivative $-\dot{\gamma}_r$ of the phase of the
wave function in Eq.\ \eqref{eq:grdot_1}
\begin{equation}
 \varepsilon_\pm^{\mathrm{(var)}} = -\dot\gamma_r
 = -\frac{4}{9\pi} \frac{5\pm 4 \sqrt{1+\frac{8 a}{3 \pi }}}
 {\left(1\pm\sqrt{1+\frac{8 a}{3 \pi }}\right)^2} \; .
\label{eq:eps_pm}
\end{equation}
The chemical potentials of the two
solutions \eqref{eq:eps_pm} are drawn in Fig.\ \ref{fig:stationary}. 
The tangent bifurcation behavior of the chemical potential at the 
critical scattering length $a_\mathrm{cr}=-3\pi/8=-1.1780$ is clearly visible.
The branch with the lower chemical potential in Fig.\ \ref{fig:stationary} 
turns out to have higher mean-field energy than the other branch.
Therefore the plus sign refers to the ground state, and the minus sign
to the collectively excited state.  
\begin{figure}[tb]
  \includegraphics[width=\columnwidth]{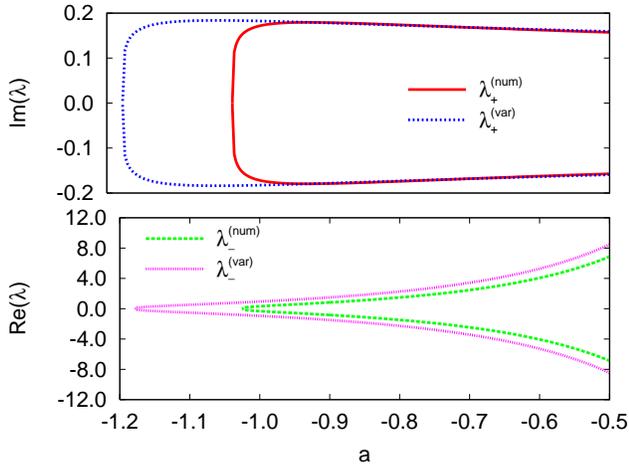}
  \caption{\label{fig:eigenvalues}(Color online) Eigenvalues of the
    linearization of the extended Gross-Pitaevskii system
    \eqref{eq:scaled_extended_GP} for the two stationary solutions emerging
    at the tangent bifurcation. Both the variational (var) and the numerically
    exact (num) solutions are shown. The two eigenvalues $\lambda_+$ of the
    stationary ground state (cf.\ Ref.\ \cite{Gio01b}) are purely
    imaginary demonstrating that the state is an elliptic fixed point. In
    contrast to this finding, the two eigenvalues $\lambda_-$ of the nodeless
    excited state are purely real. One is positive, the other is negative. The
    state is an hyperbolic fixed point. Vanishing real or imaginary parts are
    not shown.}
\end{figure}
The results obtained here via the fixed points of the time-dependent equations
of motion for the variational parameters fully agree with the results
derived in Refs.\ \cite{ODe00,Pap07} applying a time-\emph{in}dependent
variational approach to extremize the mean-field energy, and which were used
to compare with numerically exact solutions of the stationary extended
Gross-Pitaevskii equation in Refs.\ \cite{Pap07,Car07b}. 

The linearization of the equations of motion \eqref{eq:Adot_1} around the
stationary solutions $\hat{A}_r$, $\hat{A}_i$ are given by
\begin{subequations}
  \label{eq:var_lin_matrix}
  \begin{align}
    \delta \dot{A}_{r \pm} &= \pm \frac{8}{9\pi}\frac{\sqrt{1+\frac{8 a}{3\pi}}}
    {\left ( \sqrt{1+\frac{8 a}{3\pi}} \pm 1 \right )^2} \delta A_{i \pm} 
    \; ,\\
    \delta \dot{A}_{i \pm} &=  -\frac{32}{9\pi} \frac{1}{\left( \sqrt{1
          +\frac{8 a}{3\pi}} \pm 1 \right )^2} \delta A_{r \pm} \; .
  \end{align}
\end{subequations}
The eigenmodes with the eigenvalues $\lambda^{\mathrm{(var)}}$ of
Eqs.\ \eqref{eq:var_lin_matrix} are calculated with  the usual ansatz
$\delta A_{r,i}(t) = \delta A_{r,i}^{(0)} e^{\lambda^{\mathrm{(var)}} t}$. For
the stationary ground state we find the two eigenvalues
\begin{equation}
  \lambda_+^{\mathrm{(var)}} = \pm \frac{16 i}{9\pi}
  \frac{\sqrt[4]{1+\frac{8 a}{3\pi}}}{\left ( \sqrt{1+\frac{8 a}{3\pi}} 
      + 1 \right )^2} \; ,
  \label{eq:ev_stable_var}
\end{equation}
which are always imaginary for $a > -3\pi/8$ (i.e., above the bifurcation
point) and correspond to the monopole excitation mode considered in
Ref.\ \cite{Gio01b}. They describe an \emph{elliptic fixed point}, which is
stable. Furthermore, we can also give the eigenvalues of the collectively
excited state,
\begin{equation}
  \lambda_-^{\mathrm{(var)}} = \pm \frac{16}{9\pi}
  \frac{\sqrt[4]{1+\frac{8 a}{3\pi}}}{\left ( \sqrt{1+\frac{8 a}{3\pi}} 
      - 1 \right )^2} \; ,
  \label{eq:ev_unstable_var}
\end{equation}
which are always real for negative scattering lengths $a > -3\pi/8$,
and one eigenvalue is positive.  Hence, the eigenstate is unstable and
belongs to a \emph{hyperbolic fixed point}.

\subsection{Stability of the numerically exact solutions}

Numerically exact stationary eigenstates of the Gross-Pitaevskii equation
\eqref{eq:extended_GP} which are the counterpart of the variational
solutions described above have been calculated in Refs.\ \cite{Pap07,Car07b}.
Here, we investigate the stability of these states and compare the results with
the eigenvalues \eqref{eq:ev_stable_var} and \eqref{eq:ev_unstable_var}. For
the stability analysis, we extend the procedure applied by Huepe et al.\
\cite{Hue03} for condensates without $1/r$ interaction in a harmonic trap
to the scaled integro-differential equation  \eqref{eq:scaled_extended_GP}.
The extended Gross-Pitaevskii equation is linearized with the Fr\'echet
derivative, once the wave function has been decomposed in its real
$\psi^R(\bm{r},t)$ and imaginary $\psi^I(\bm{r},t)$ part. The derivative is
evaluated at the real-valued stationary states $\hat{\psi}_\pm(\bm{r})$ 
found in Ref.\ \cite{Pap07}, which leads to two coupled equations, viz.
\begin{widetext}
  \begin{subequations}
    \label{eq:num_lin_GP}
    \begin{align}
      \frac{\partial}{\partial t}\delta\psi^R &= \left ( -\Delta
        -\varepsilon + 8\pi a \hat{\psi}_\pm(\bm{r})^2 - 2  \int 
        d\bm{r}'\, \frac{\hat{\psi}_\pm(\bm{r}')^2}{|\bm{r}-\bm{r}'|}
      \right ) \delta\psi^I(\bm{r},t)\; , \\
      \frac{\partial}{\partial t}\delta\psi^I &= \left ( -\Delta
        -\varepsilon + 24\pi  a \hat{\psi}_\pm(\bm{r})^2 - 2 
        \int d^3 \bm{r}'\, \frac{\hat{\psi}_\pm(\bm{r}')^2}
        {|\bm{r}-\bm{r}'|} \right ) \delta\psi^R(\bm{r},t)
      + 4  \hat{\psi}_\pm(\bm{r}) \int d^3 \bm{r}'
      \frac{\hat{\psi}_\pm(\bm{r}')\,\delta\psi^R(\bm{r}',t)}
      {|\bm{r}-\bm{r}'|} \; .
    \end{align}
  \end{subequations}
\end{widetext}
For the calculation of the eigenmodes we take the perturbations
in the form
$\delta\psi^R(\bm{r},t) = \delta\psi^R_0(\bm{r})
\mathrm{e}^{\lambda^{\mathrm{(num)}} t}$, and $\delta\psi^I(\bm{r},t) =
\delta\psi^I_0(\bm{r})\mathrm{e}^{\lambda^{\mathrm{(num)}} t}$. Note
that $\delta\psi^R_0(\bm{r})$ and $\delta\psi^I_0(\bm{r})$ are
not always real-valued. As solutions of the linearized equations
\eqref{eq:num_lin_GP} they can become complex wave functions.

The numerical solution of the linearized wave equations is done
with the method described in Ref.\ \cite{Car07b}. Using the ``linearized
potential''
\begin{align}
  U_1(\bm{r}) = 4 \int d^3 \bm{r}' \frac{\hat{\psi}_\pm(\bm{r}')\,
    \delta\psi^R(\bm{r}',t)}{|\bm{r}-\bm{r}'|}\; ,
\end{align}
we can transform \eqref{eq:num_lin_GP} into a system of three second order
ordinary differential equations, which can be solved together with the
stationary extended Gross-Pitaevskii equation. In order to be in a 
position to compare with  the results of the
variational approach we assume radially symmetric wave functions. 
The ordinary differential equations, which have to be solved, then read
\begin{widetext}
  \begin{subequations}
    \label{eq:num_lin_system}
    \begin{align}
      \lambda \delta\psi^R_0(r) &=  -\delta\psi^I_0(r)'' 
      - \frac{2}{r} \delta\psi^I_0(r)' 
      + \left ( 8\pi a \hat{\psi}_\pm(r)^2 - U(r)\right )  \delta\psi^I_0(r)
      \; , \\
      \lambda \delta\psi^I_0(r) &=  -\delta\psi^R_0(r)'' 
      - \frac{2}{r} \delta\psi^R_0(r)' 
      + \left ( 24\pi a \hat{\psi}_\pm(r)^2 - U(r)\right )  \delta\psi^R_0(r)
      - U_1(r) \hat{\psi}_\pm(r) \; , \\
      U_1(r)'' & = - \frac{2}{r} U_1(r)' -16\pi  \,\hat{\psi}_\pm(\bm{r})\,
      \delta\psi^R(\bm{r},t) \; . 
    \end{align}
  \end{subequations}
\end{widetext}
The wave functions $\delta\psi^R_0(r)$ and $\delta\psi^I_0(r)$ are only
determined up to a complex (normalization) constant, which they have in
common. Therefore, only two real variables are required to set the initial
value $(\delta\psi^R_0(0), \delta\psi^I_0(0)) = (\cos\alpha, \sin\alpha\,
e^{i \beta})$ for the integration.
The real angles $\alpha$ and $\beta$, the complex eigenvalue $\lambda$, and 
the complex initial value $U_1^{(0)} = U_1(0)$ are the parameters which have to
be searched to find the solutions of the system \eqref{eq:num_lin_system}.
The initial conditions for the first derivatives are $(\delta\psi^{R})' =
(\delta\psi^I)' = U_{1}' = 0$. The correct solutions are found when we obtain
square integrable wave functions and  the value of $U_1^{(0)}$ initially
assumed is identical to the integral
\begin{equation}
  U_1(0) = 16\pi \int_0^\infty dr' \, r' \hat{\psi}_\pm(r')\, \delta\psi_0^R(r')
  \; ,
\end{equation}
which can be calculated once the wave function  
obtained numerically has been
determined.

As the Gross-Pitaevskii equation is a nonlinear differential equation
the correct normalization of the numerical wave functions cannot be obtained
by a multiplication with a normalization constant but is possible by
exploiting scaling properties of equations \eqref{eq:num_lin_system}.
In the form presented here, the proper normalization and scaling of the
extended Gross-Pitaevskii system is achieved by the scaling factor
\begin{equation}
  1/\nu = ||\psi||^2 = 4\pi\int_0^\infty dr\, |\psi(r)|^2 r^2
\end{equation}
and the transformation of \emph{all} of the following values (cf.\ 
\cite{Car07b}):
\begin{equation}
  (\psi,r,\varepsilon,t,a,U) \to (\nu^2\psi, \frac{r}{\nu}, \nu^2\varepsilon,
  \frac{t}{\nu^2}, \frac{a}{\nu^2},\nu^2U)\; . \label{eq:scaling_transf_1}
\end{equation}
As can be seen from the system \eqref{eq:num_lin_system}, the differential
equations are invariant if in addition to the transformations
\eqref{eq:scaling_transf_1} the eigenvalue is scaled by $\lambda \to \nu^2
\lambda$. Thus, it is possible to solve the system of differential equations
\eqref{eq:num_lin_system} with the  unscaled and non-normalized 
stationary solutions $\hat{\psi}_\pm(r)$ of the nonlinear
extended Gross-Pitaevskii equation computed numerically if a subsequent 
scaling of the eigenvalues is performed.

When the solution of the linearized system is carried out for the stationary
ground state, a pair
$\lambda_1 = -\lambda_2$ of purely imaginary
eigenvalues are found. These are plotted as a function of the scaled
scattering length $a$ in Fig.\ \ref{fig:eigenvalues}.
The same calculation  for the nodeless stationary state yields a pair of purely real eigenvalues. The
qualitative agreement between the  eigenvalues  calculated numerically
and the results obtained analytically in the variational approach is very good,
but there are significant quantitative differences. Such  behavior is
known from previous studies of the system \cite{Pap07,Car07b}.
Similar to the case discussed in Ref.\ \cite{Hue03}, there exist neutral modes
with the eigenvalue $\lambda^{\mathrm{(num)}} = 0$, and additional imaginary
eigenvalues are found for both stationary solutions. Altogether, the
numerically exact results confirm the stability analysis performed using 
the analytical eigenvalues of the variational approach.

\section{Dynamics of the condensate}
\label{sec:dynamics}
In this section we investigate the time evolution of arbitrary
wave functions. We do this again first by the variational approach and then
by exact numerical calculations. The variational calculations will reveal 
the different types of dynamics in the vicinity of the elliptic
and the hyperbolic fixed points, viz. 
oscillatory or collapsing solutions. Furthermore, the collapse of the
condensate for $a<a_\mathrm{cr}$ can be followed as a function of time. 
The numerically exact approach will confirm these findings and, in addition,
exhibit a larger variety of qualitatively different dynamical behavior of the
condensate.

\subsection{Variational approach}
We solve the two ordinary differential equations \eqref{eq:Ardot_1} and
\eqref{eq:Aidot_1} for various scaled scattering lengths $a$
above and below the critical value $a_\mathrm{cr}=-3\pi/8=-1.1780$. 
Phase portraits of the dynamics can be obtained by plotting the imaginary
part as a function of the real part of the time-dependent trajectories
$A(t)=A_r(t)+iA_i(t)$.
\begin{figure}[tb]
  \includegraphics[width=\columnwidth]{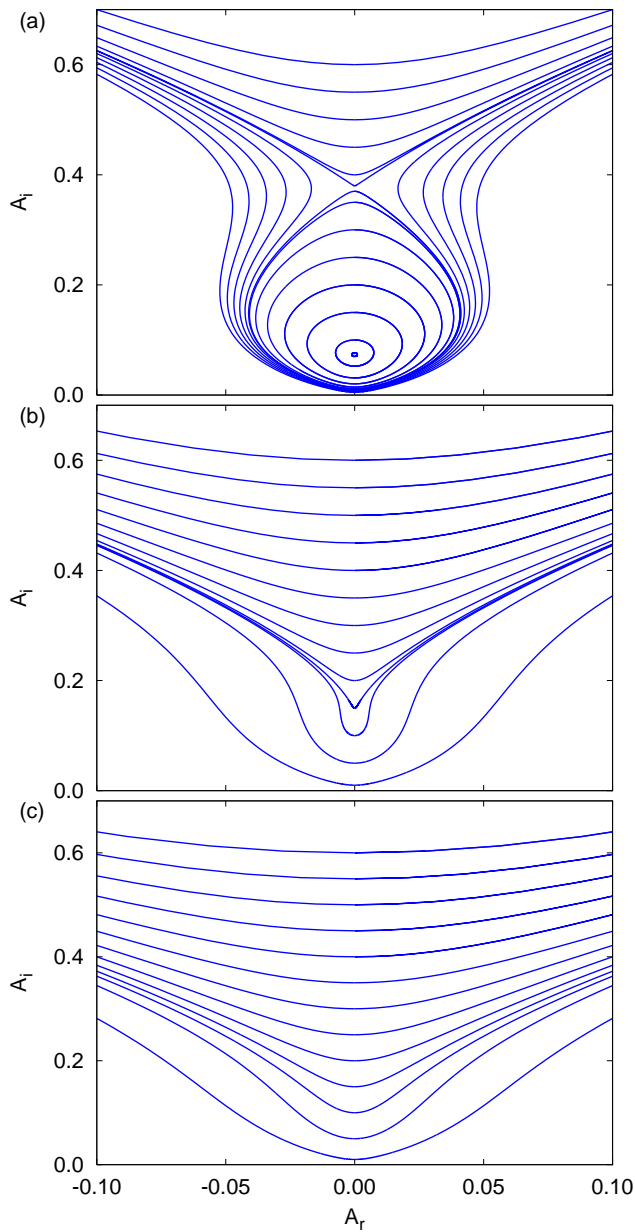}
  \caption{\label{fig:phase_portrait_A}(Color online)
    Phase portraits of the dynamics obtained from time-dependent trajectories
    of the complex function $A(t)$. (a) For $a=-1$ the two stationary solutions
    appear as fixed points. (b) They coalesce for $a=-1.18$. (c) For $a=-1.3$,
    below the bifurcation point, there are no stationary solutions.}
\end{figure}
The phase portraits of trajectories with different initial conditions
are shown in Fig.\ \ref{fig:phase_portrait_A} for three values 
of the scaled scattering length, one above the bifurcation point 
($a=-1$, Fig.\ \ref{fig:phase_portrait_A}(a)), one near the bifurcation point
($a=-1.18$, Fig.\ \ref{fig:phase_portrait_A}(b)), and one below the
bifurcation point ($a=-1.3$, Fig.\ \ref{fig:phase_portrait_A}(c)). 
For $a>a_\mathrm{cr}$ the elliptic and the hyperbolic fixed points 
are clearly recognizable in Fig.\ \ref{fig:phase_portrait_A}(a), they
correspond to the stationary eigenstates.
The two fixed points coalesce at the critical scattering length $a_\mathrm{cr}$
(see Fig.\ \ref{fig:phase_portrait_A}(b)), and disappear for $a<a_\mathrm{cr}$
(Fig.\ \ref{fig:phase_portrait_A}(c)) implying that no longer stationary
eigenstates exist.

For the physical interpretation of the phase portraits it is useful to
note that the width $\sqrt{\langle r^2\rangle(t)}$ of the condensate is
related to $A_i(t)$ via
\begin{equation}
   \langle r^2\rangle(t)
 = \frac{\langle\psi|r^2|\psi\rangle}{\langle\psi|\psi\rangle}
 = \frac{3}{4 A_i(t)} \; . \label{eq:width}
\end{equation}
Thus, in the stable region surrounding the elliptic fixed point the width
of the condensate oscillates periodically.
\begin{figure}[tb]
  \includegraphics[width=\columnwidth]{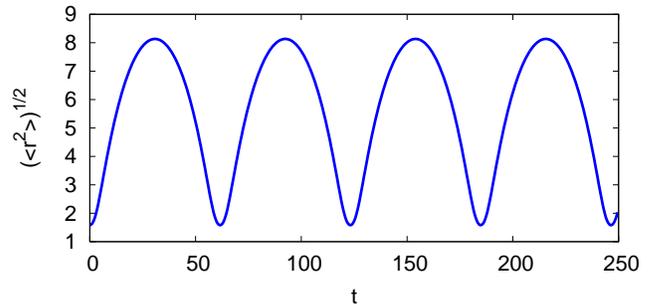}
  \caption{\label{fig:gauss_osc}(Color online)
    Periodically oscillating condensate for the scattering length $a=-1$ and
    the initial condition $A_i(0)=0.3$.}
\end{figure}
This is illustrated in Fig.\ \ref{fig:gauss_osc} for a condensate with
scattering length $a=-1$ and initial condition $A_i(0)=0.3$.

In the unstable regions $A_i(t)$ increases to infinity, which means the
collapse of the condensate.
\begin{figure}[tb]
  \includegraphics[width=\columnwidth]{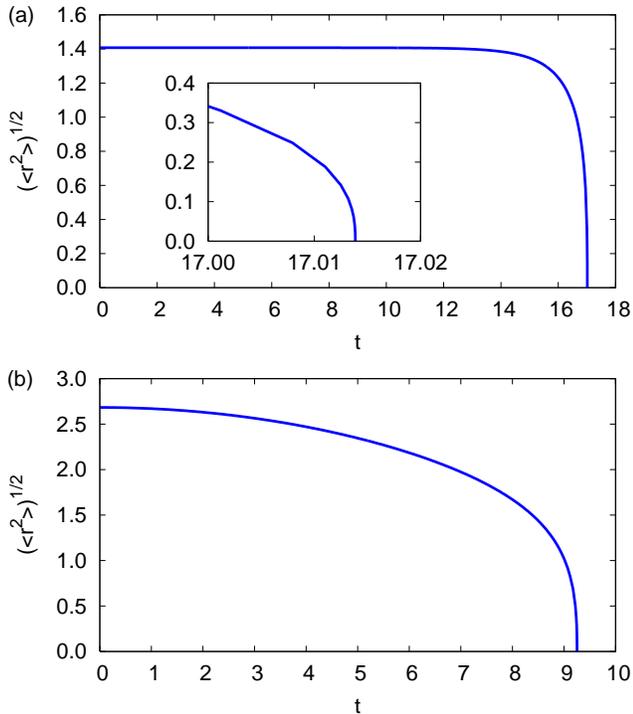}
  \caption{\label{fig:collapse_unstable}(Color online)
 Collapse of the unstable eigenstate.
 (a) Scaled scattering length $a=-1$ and initial condition close
 to the hyperbolic fixed point at $A_r=0$, $A_i=0.3787$.  The collapse
 after finite time is clearly visible in the inset,
 but depends sensitively on the initial conditions (see text).
 (b) Collapse of the non-stationary state at $a=-1.3$ with initial condition
 $A_i(0)=\frac{1}{6a}+\frac{\pi}{8a^2}=0.10416$.}
\end{figure}
In Fig.\ \ref{fig:collapse_unstable}(a) the width $\sqrt{\langle r^2\rangle}$
is shown for a condensate with scaled scattering length $a=-1$ and initial
condition close to the hyperbolic fixed point, given by
Eq.\ \eqref{eq:fixed_points_Ar}, \eqref{eq:fixed_points_Ai} at $A_r=0$,
$A_i=0.3787$. 
The width first stays approximately constant, as is to be expected
in the vicinity of the unstable stationary state, however, as soon as the
decrease of the width becomes noticeable, the complete collapse to zero width 
occurs within finite time.
This behavior demonstrates the existence of a collapse induced
by the attractive atom-atom interactions. In a realistic experimental
situation further mechanisms, which go beyond the scope of this paper,
have to be taken into account. In particular, the contraction of the
wave function amplifies density-dependent inelastic collisions which result
in a loss of particles \cite{Kag98,Don01} and change the time evolution.
Note that the calculations presented here assume a constant scaled 
scattering length (cf.\ Eq.\ \eqref{eq:scaling}).
Of course, for initial conditions close to the hyperbolic fixed point
the evolution of the collapse depends sensitively on the initial deviation
from the fixed point, i.e., the closer the initial conditions approach the
hyperbolic fixed point the longer it takes before the collapse sets in.

At scattering lengths $a<a_\mathrm{cr}$ the square root in
Eq.\ \eqref{eq:fixed_points_Ai} becomes imaginary which means that no fixed
points and thus no stationary states can exist.
The condensate collapses for all initial conditions $A(0)$ of the wave
function.
This is illustrated for $a=-1.3$ in Fig.\ \ref{fig:collapse_unstable}(b) with
the initial condition of the ``least'' unstable non-stationary state, 
$A_i(0) = \frac{1}{6a} + \frac{\pi}{8a^2} = 0.10416$ 
(cf.\ Eq.\ \eqref{eq:fixed_points_Ai}). Contrary to the situation shown in
Fig.\ \ref{fig:collapse_unstable}(a) the decrease of the condensate width
starts immediately without any plateau. Ignoring a loss of
particles we find that the width vanishes after the finite time $T_c= 9.2522$.

The variational approach with complex parameters $A(t)$ and
$\gamma(t)$ in Eq.\ \eqref{eq:psi} ensures that the mean-field energy
\begin{equation}
  E = \frac{3 \left(A_i^2+A_r^2\right)}{A_i}+
  \frac{2 \sqrt{A_i} (2 a A_i-1)}{\sqrt{\pi }} 
\end{equation}
is conserved. Thus, the phase portraits in Fig. \ref{fig:phase_portrait_A}
can be calculated alternatively as equipotential lines $E=\mathrm{const}$
instead of integrating the equations of motion \eqref{eq:Adot_1}.

In fact, the equations of motion obtained from the 
time-dependent variational principle can be brought into Hamiltonian form
if the variational
parameters $A_r,A_i$ are replaced with two other dynamical quantities,
of which one is assigned to be the momentum and the other  the
coordinate variable. 
Such adequate canonical variables are \cite{Bro89}
\begin{subequations}
  \begin{align}
    q &= \frac{1}{2}\sqrt{\frac{3}{A_i}} = \sqrt{\langle r^2 \rangle}\,, \\
    p &= A_r\sqrt{\frac{3}{A_i}}\,,
  \end{align}
\end{subequations}
and in this set the mean-field energy reads
\begin{equation}
 E = H(q,p) = T+V = p^2+\frac{9}{4q^2}
 +\frac{3 \sqrt{3}a}{2\sqrt{\pi}q^3} -\frac{\sqrt{3}}{\sqrt{\pi}q} 
 \label{eq:mfe_class}
\end{equation}
with the decomposition into a ``kinetic'' part $T$ depending on the ``momentum''
$p$ and a ``potential'' part $V$ depending only on the ``coordinate'' $q$.
Note that $q$ has the physical meaning of the square root of the radius of the
condensate, according to equation \eqref{eq:width}. 
If the mean-field energy \eqref{eq:mfe_class} is identified with a
Hamiltonian, it describes the one-dimensional motion of a particle in the
potential $V(q)$ obeying Hamilton's equations:
\begin{subequations}
  \begin{align}
    \dot q &= \frac{\partial H}{\partial p} = 2 p \; ,\label{eq:q_hamil} \\
    \dot p &= -\frac{\partial H}{\partial q}
    =  \frac{9}{2q^3}+\sqrt{\frac{3}{\pi }} \frac{9a}{2q^4}
    -\sqrt{\frac{3}{\pi }}\frac{1}{q^2} \; .  \label{eq:p_hamil}
  \end{align}
\end{subequations}
Of course the backward substitution $A_r=\frac{p}{2q}$, $A_i=\frac{3}{4q^2}$
together with \eqref{eq:q_hamil} and \eqref{eq:p_hamil} yields the
same equations of motion for $A_r$ and $A_i$ as obtained from the TDVP.
Conversely, if the trial wave function \eqref{eq:psi} had been parametrized
by $q$, $p$, $\gamma$, the TDVP would have yielded their equations of
motion \eqref{eq:q_hamil} and \eqref{eq:p_hamil}. 

\begin{figure}[tb]
  \includegraphics[width=\columnwidth]{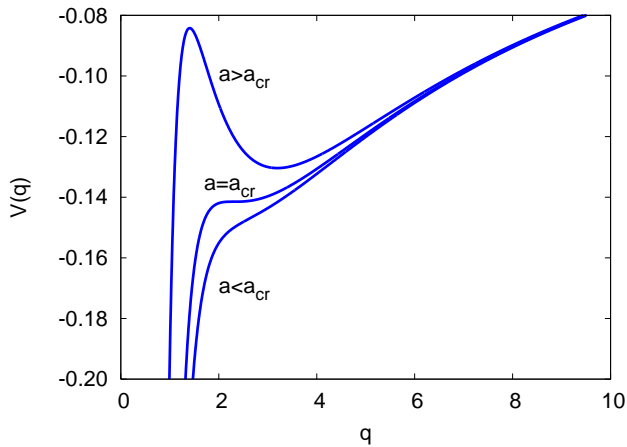}
  \caption{\label{fig:potential}(Color online) ``Potential'' part of the
    mean-field energy for different scattering lengths. This potential part of
    the mean field energy \eqref{eq:mfe_class} agrees with the complete
    mean-field energy for \emph{stationary states} of Ref.\ \cite{ODe00} (see
    text). The motion of the Gaussian wave function is interpreted as the 
    one-dimensional motion of a classical particle in the potential $V(q)$.}
\end{figure}

The ``potential'' part $V(q)$ of the mean-field energy \eqref{eq:mfe_class}
is plotted in Fig. \ref{fig:potential} as a function of the ``position''
variable $q$ for different scattering lengths below, at, and above the critical
scattering length $a_{cr}$. It agrees with the mean-field energy plotted in
Ref.\ \cite{ODe00}, calculated using  a real-valued spherically symmetric 
(stationary) Gaussian trial wave function.
In our approach the ``kinetic'' term $p^2$ in Eq.\
\eqref{eq:mfe_class} appears additionally in the mean-field energy because of
the complex ansatz \eqref{eq:psi}. In other words, the full mean-field
energy of Ref.\ \cite{ODe00} corresponds to our potential part $V$, in
which the condensate moves like a classical particle.

Above $a_\mathrm{cr}$ the potential possesses
two stationary points, a stable one at the minimum  and an 
unstable one at the maximum. At $a_\mathrm{cr}$ the bifurcation
takes place, i.e., the two extrema coincide and there is only a saddle point
of the potential. For $a < a_{cr}$ the potential has no stationary points.
For $a > a_\mathrm{cr}$ the motion of the condensate is stable as 
long as the mean-field energy lies below the  
maximum of the peak of the potential and if it
is located close to the local minimum on the right-hand side of the unstable
fixed point.
The one-dimensional motion is periodic between two turning points.  
If the energy is increased above the energy of the unstable fixed point,
only one turning point remains, and the condensate collapses
when $q$ approaches zero.

Expanding the potential about the minimum and maximum leads to the 
frequencies  \eqref{eq:ev_stable_var} of oscillations of the ground state and
the decay rates \eqref{eq:ev_unstable_var} of the collectively excited state.
\begin{figure}[tb]
  \includegraphics[width=\columnwidth]{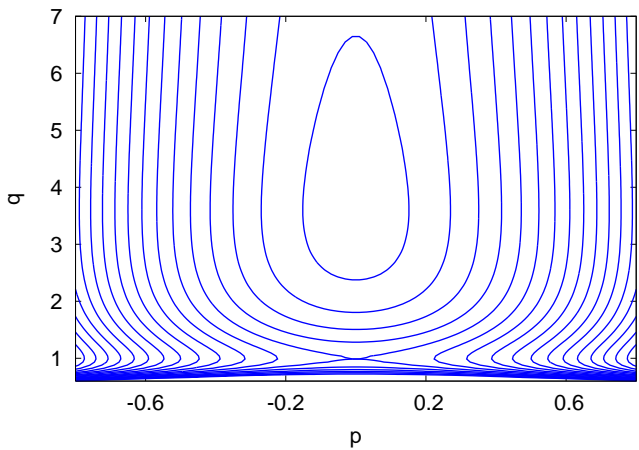}
  \caption{\label{fig:phase_portrait_pq}(Color online)
    Phase portrait of the mean-field energy formulated in the canonical
    variables $q$ and $p$ for $a=-0.8$.}
\end{figure}
The phase space portrait of the dynamics in $(p,q)$-variables is presented
in Fig.\ \ref{fig:phase_portrait_pq} for the scattering length $a=-0.8$. The
equipotential lines which asymptotically approach the $q=0$ axis represent
the collapse of the condensate.

\subsection{Exact time-dependent calculations with the split-operator method}
The split-operator method \cite{Fei82} assumes the decomposition of the
Hamiltonian $H=T+V$, and makes use of the short-time approximation for the 
time evolution operator
\begin{equation}
e^{-i\tau(T+V)} =  e^{-i(\tau/2)T}e^{-i\tau V}e^{-i(\tau/2) T} +O(\tau^3) \; .
\end{equation}
The kinetic part of the evolution operator is applied to the wave function in
momentum space, the potential part is applied to the wave function in position
space representation. 
The method is numerically 
especially efficient when a fast Fourier transform is used for  
the transition from momentum to position space representation 
and backwards.
For the nonlinear Gross-Pitaevskii equation \eqref{eq:scaled_extended_GP} the
potential part of the time evolution operator needs some further investigation
in comparison with the linear Schr\"odinger equation.
Although the scattering term $V_c$ belongs to the nonlinear part of the
Gross-Pitaevskii equation, it can be treated like a conventional potential of
a linear Schr\"odinger equation and presents no additional difficulty.
The $1/r$ interaction potential $V_u$ however requires the additional solution
of the integral \eqref{eq:integral_Vu} after every time step of integration.
This integral is computed using the convolution theorem, i.e.,
\begin{equation}
  \mathcal{F}\{V_u(r,t)\} = \mathcal{F}\{\frac{1}{r}\}\mathcal{F}
  \{| \psi(r,t)|^2\} \; ,
\end{equation}
where the Fourier transform of $1/r$ is performed analytically to give
\begin{equation}
\mathcal{F}\{\frac{1}{r}\}=\frac{4 \pi}{p^2}\; .
\end{equation}
Altogether two additional fast Fourier transforms are necessary per time step 
and we obtain for the $1/r$ interaction term
\begin{gather}
V_u(r,t)  = 
 -\frac{16}{r}\int_0^\infty dp \frac{\sin pr}{p^2}\int_0^\infty
dr' r'| \psi(r',t)|^2\sin pr' \; .
\end{gather}

The fast Fourier transforms are performed on equidistant grids
with 1024 or 2048 points. The numerical convergence is checked
by monitoring that the wave function vanishes at the $r_{\mathrm{max}}$, 
$p_{\mathrm{max}}$ border of the grid in
position as well as in momentum space. Due to the widening of the wave
function in some computations the necessary size of the grid depends on the
desired propagation time. A criterion of convergence for the time step of
the split-operator method is that the mean-field energy is a constant of 
motion. In particular for the computations in which the wave function is
very close to the hyperbolic fixed point the time step must be chosen small
for convergence. Specifically, for the computations in the vicinity of the
hyperbolic fixed point we use $\Delta t = 0.0001$ whereas for computations
far away from the unstable stationary state a time step of $\Delta t = 0.01$ is
sufficient.

In the following paragraphs, the dynamics of the condensate is
investigated for regions initially close to the two stationary states.
In particular, we present the evolution of those initial states which are
obtained by deforming the stable and the unstable stationary state by  
\begin{equation}
  \psi \rightarrow \psi\cdot f,\quad r \rightarrow r/f^{2/3} \label{eq:def}
\end{equation}
with a stretching factor $f$, i.e., for $f=1$ the stable and the unstable
stationary state, respectively, are obtained. This choice of perturbation
leaves the norm of the wave function unchanged.

\begin{figure}[tb]
  \includegraphics[width=\columnwidth]{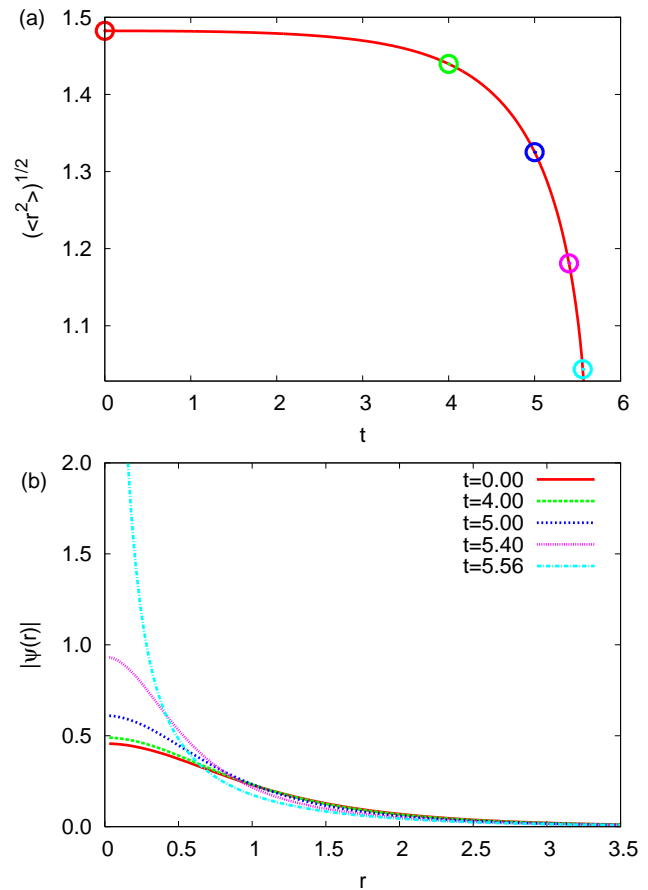}
  \caption{\label{fig:collapse_num}(Color online) (a) Width of a slightly
    perturbed stationary state ($a=-0.85$, $f=1.001$) as a function of
    time.
    (b) Wave functions of the state for selected times marked by circles in (a).
    The collapse of the condensate 
    is obvious from both plots.}
\end{figure}

We start with the investigation of the dynamics of wave functions close to
the unstable stationary state. In Fig.\ \ref{fig:collapse_num}(a) the square
root of the width \eqref{eq:width} of the condensate is plotted as a function
of time. The evolution presented in Fig.\ \ref{fig:collapse_num}(a) is
computed at the scaled negative scattering length $a=-0.85$ with a deformation
factor $f=1.001$. The wave function itself is plotted in Fig.\
\ref{fig:collapse_num}(b) as a function of the radial coordinate $r$ at
different times. The initial wave function is given by the solid red line. The
condensate stays nearly stationary at the beginning for times up to
$t\approx 4$, i.e., the dashed green line in Fig.\ \ref{fig:collapse_num}(b)
representing the wave function at $t=4$ nearly coincides with the initial
state, and the width (Fig.\ \ref{fig:collapse_num}(a)) has only slightly
decreased from $\sqrt{\langle r^2\rangle} = 1.48$ at $t=0$ to
$\sqrt{\langle r^2\rangle} = 1.44$ at $t=4.0$.     
With increasing time and the farther away the wave function has moved from the
stationary state, the shrinking of the width of the wave function is
accelerated. As already predicted by the variational computation
this leads to a collapse of the condensate at finite time.
Both Figs.\ \ref{fig:collapse_num}(a) and \ref{fig:collapse_num}(b) show that
the width of the condensate tends to zero in position space when the collapse
time $T_c$ is approached.
Conversely, in momentum space the wave function becomes arbitrarily wide
close to $T_c$. Thus, the size of the grid in momentum space is the numerically
limiting factor for the propagation with the split-operator method. Choosing
a large grid in momentum space allows for integrating arbitrarily close to
$T_c$. 
 
\begin{figure}[tb]
  \includegraphics[width=\columnwidth]{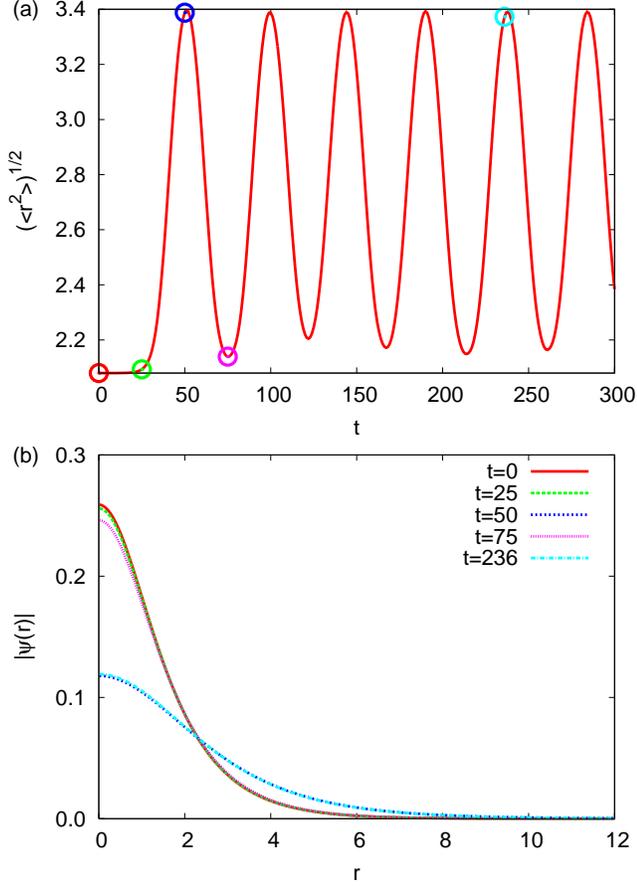}
  \caption{\label{fig:osc_num}(Color online) Evolution of the  unstable  
    ``stationary'' state ($f=1$) at $a=-1.0$. (a) Due to numerical deviations
    an oscillation of the width starts for times larger than $t\approx 25$.
    (b) Corresponding wave functions for the times marked in (a).
}
\end{figure}

According to the variational results there exist periodic solutions in
the vicinity of the unstable stationary state, and, indeed, similar behavior
is found in the numerically exact computations. In Fig.\ \ref{fig:osc_num}(a)
the width of a quasi-periodically oscillating condensate at $a=-1.0$ is shown.
The associated wave function is plotted in Fig.\ \ref{fig:osc_num}(b) for
different times.    
It can be seen that at the  times where the root-mean-square extension
of the condensate goes through a maximum, or minimum, there is also a good 
agreement of the respective wave functions. This shows that the wave functions
indeed oscillate quasi-periodically.
The calculation is started with the wave function of the unstable stationary
state ($f=1$). Because of numerical deviations, the solution begins to
oscillate for times larger than $t\approx 25$. In contrast to the variational
result, the oscillation of the condensate is not strictly periodic here.

\begin{figure}[tb]
  \includegraphics[width=\columnwidth]{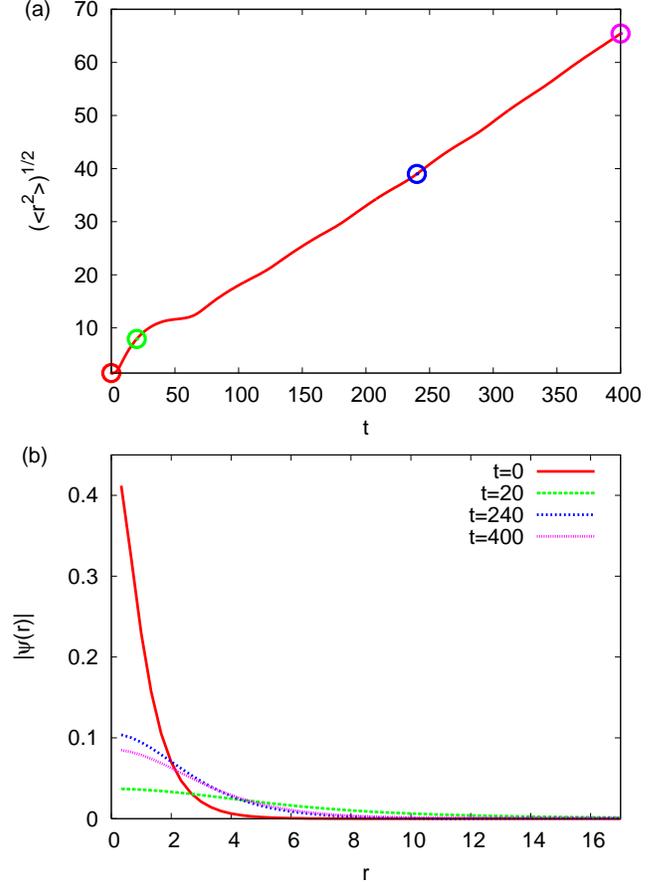}
  \caption{\label{fig:increasing_width}(Color online) (a) The width of the
    condensate, initially in the vicinity of the unstable fixed point with
    $f=0.99$ and $a=-0.85$ increases linearly with time for $t \gtrsim 65$.
    (b) Wave functions of the same state for different times
marked by circles in (a). The wave
    functions at $t=240$ and $t=400$ seem to be sharper than the wave
    function at $t=20$. Note that a larger step size on the grid 
    in position space than in Figs.\ \ref{fig:collapse_num} and
    \ref{fig:osc_num} was used. Therefore there is a visible distance
    between $r=0$ and the first point on the grid.}
\end{figure}

\begin{figure}[tb]
  \includegraphics[width=\columnwidth]{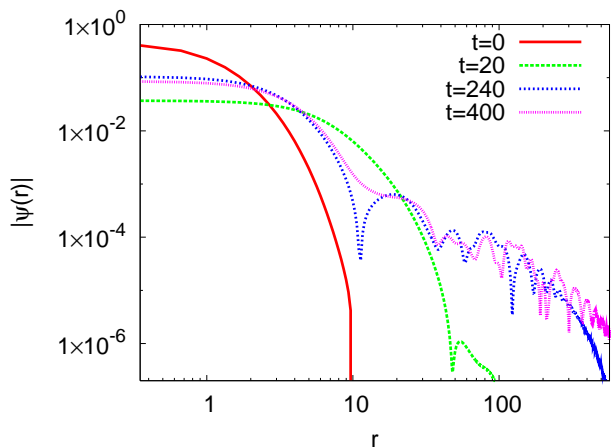}
  \caption{\label{fig:increasing_width_log}(Color online)
    Double logarithmic plot of the wave functions presented in Fig.\ 
    \ref{fig:increasing_width}(b). The long range tail occurring with
    increasing propagation times and leading to a monotonically increasing
    width in Fig.\ \ref{fig:increasing_width}(a) becomes visible.}
\end{figure}

The dynamics of the condensate presented so far does not qualitatively differ
from what is predicted by the variational calculation. In Fig.\
\ref{fig:increasing_width}(a), however, the situation is encountered where the
width increases monotonically with time. Here we have chosen a deformation of
$f=0.99$ at the scaled scattering length $a=-0.85$. 
For short times there is again a plateau where the width is nearly constant
as in Fig.\ \ref{fig:osc_num}(a). The plateau is shorter than that in
Fig.\ \ref{fig:osc_num}(a) because the initial wave function of Fig.\
\ref{fig:increasing_width}(b) differs more from the stationary state than the
initial state associated with Fig.\ \ref{fig:osc_num}(a). For times $t
\gtrsim 65$, however, the width increases linearly with time.   
Obviously Fig.\ \ref{fig:increasing_width}(b) shows a broadening of the wave
function for times up to $t \approx 20$.  For longer times the main peak of the
wave function located at the origin does not broaden monotonically with time
as might be concluded from Fig.\ \ref{fig:increasing_width}(a). Indeed, the
wave functions at $t=240$ and $t=400$ seem to be even sharper than the wave
function at $t=20$ in Fig.\ \ref{fig:increasing_width}(b). 
This apparent contradiction is resolved, if the wave
functions are plotted on a broader range in position space and on a logarithmic
scale, which is presented in Fig.\ \ref{fig:increasing_width_log}. Here it is
clearly visible that the wave functions have more than one peak and their
width \emph{does} increase for larger times. However, the amplitude of the 
run-away parts is very small compared to the first maximum at the origin. For
obtaining the accurate propagation it is therefore necessary to choose a large
grid in position space to make the wave function vanish on the border, in
particular for long propagation times. The usual method to prevent the
wave function from running to the border of the grid and being reflected there,
namely to introduce an absorbing complex potential, is not applicable for the
nonlinear Gross-Pitaevskii system, in which an absorption of the wave function
alters the potential.  
\begin{figure}[tb]
  \includegraphics[width=\columnwidth]{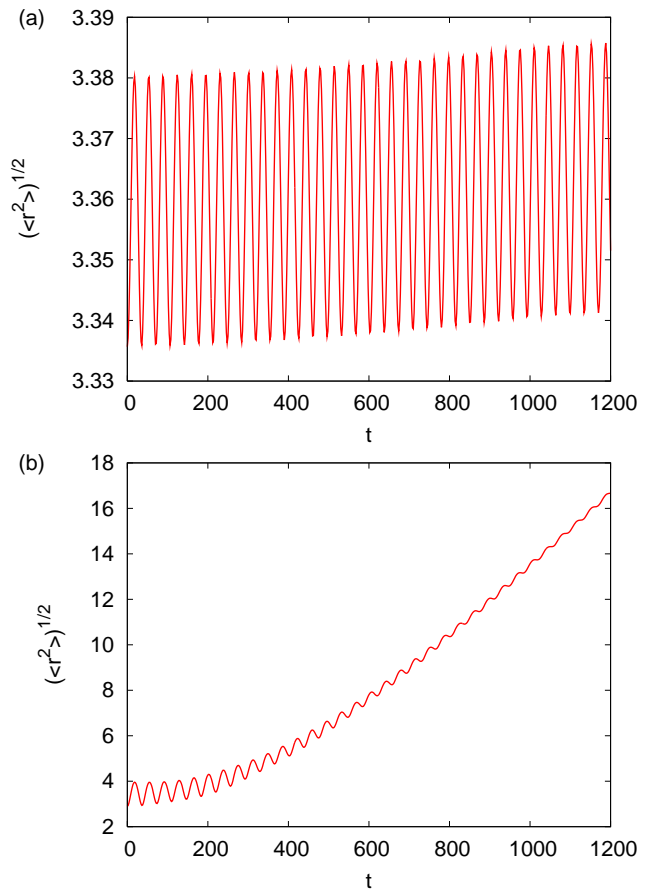}
  \caption{\label{fig:osc_expanding}(Color online)
    (a) Quasi periodically oscillating condensate at a scaled scattering
    length of $a=-0.85$ and $f=1.01$.
    (b) An initially quasi periodically oscillating condensate at $a=-0.85$,
    $f=1.25$ turns into expanding dynamics after long time.}
\end{figure}

In the vicinity of the stable stationary state the condensate exhibits an
oscillatory behavior shown in Fig.\ \ref{fig:osc_expanding}(a). The parameter
values for this computation are a stretching of $f=1.01$ of the stable
stationary state at $a=-0.85$. The amplitude of the oscillation is very small
compared to the oscillation in Fig.\ \ref{fig:osc_num}(b) and ranges from about 
$\sqrt{\langle r^2\rangle} = 3.336$ to $\sqrt{\langle r^2\rangle} = 3.385$.
This observation is in correspondence with the variational result in
Fig.\ \ref{fig:phase_portrait_A}(a) where small deviations from the stable
fixed point lead to small oscillations around the fixed point along the
equipotential lines whereas small deviations from the unstable fixed point
may lead to oscillations with a large amplitude. 
However, with increasing distance from the stable fixed point the dynamics is
no longer oscillatory, but, as can be seen in Fig.\
\ref{fig:osc_expanding}(b), there is also a gradual broadening of the
condensate, which becomes more pronounced for larger deviations of the initial
wave function from the stable stationary state as is shown in Fig.\
\ref{fig:osc_expanding}(b), in which the time evolution of the width for
$f=1.25$ is plotted. 
The scattering length is set to $a=-0.85$ again. Here, the oscillatory
motion at the beginning changes to a mainly expanding motion with minor
modulations.

\section{Conclusion}
\label{sec:disc}

We have investigated the time-dependent extended Gross-Pitaevskii equation for
self-bound Bose-Einstein condensates with attractive $1/r$ interaction. The two
stationary states of the system, which emerge in a tangent bifurcation,
and were found with time-independent calculations in Refs.\ \cite{ODe00,Pap07},
were identified as an elliptic and a hyperbolic fixed point of the dynamical
system. A time-dependent variational approach opened the possibility to
obtain analytical results for the eigenvalues of the linearized equations
of motion, which is a feature of the self-trapped condensate with
$1/r$ interaction. Numerically exact
calculations confirmed the variational findings.

The dynamics of the system was analyzed with the time-dependent variational
approach and numerically exact computations with the split-operator method.
For scaled scattering lengths larger than the critical value an oscillatory 
behavior of the condensate was found in the vicinity of the stationary 
solutions. It was shown that it is possible to introduce canonical 
variables and interpret the dynamics of the width of the condensate as the
motion of a classical particle in a potential. Furthermore, there are unstable
regions of the phase space, in which a collapse of the condensate occurs.
For scaled scattering lengths below the critical value the condensate is
always collapsing. The numerically exact method revealed solutions with
a continuously expanding motion of the width.

The self-bound spherically symmetric Bose-Einstein condensates with gravity-like
interaction investigated in this paper are unique in so far as  variational
calculations to a large extent can be performed analytically. The results
obtained in this way can serve as a useful guide to numerical calculations,
and to the exploration of the nonlinear dynamic properties of more general
Bose-Einstein systems. The appearance of stable elliptic islands, with
elliptic fixed points, unstable regions, with hyperbolic fixed points, and of
separatrices dividing the regions, are typical signatures of  autonomous
Hamiltonian systems. Therefore they will also be present if a symmetric trap
potential is added to the contact and gravity-like interaction.

In anisotropic condensates the variational approach will lead  to 
autonomous Hamiltonian dynamics with more than one degree of freedom. Therefore
in addition to the signatures discussed in this paper
signatures of chaos can appear. In fact,  a recent variational investigation 
of Bose-Einstein
condensates with a dipole-dipole interaction \cite{Wag08} has revealed
that,  as the mean-field energy is increased from its ground-state value,  
chaotic regions emerge in phase space which coexist 
with regular islands, corresponding to irregularly fluctuating
or quasi-periodically oscillating states of the condensates, respectively. 
The investigations presented in this paper have paved the way for studies of
this sort in realistic Bose-Einstein condensates with long-range interactions.

\begin{acknowledgments}
  This work was supported by Deutsche Forschungsgemeinschaft. H.C. 
  is grateful for support from the Landesgraduiertenf\"orderung of
  the Land Baden-W\"urttemberg.
\end{acknowledgments}

%\bibliography{paper.bib,bec_bank_1}

\begin{thebibliography}{27}
\expandafter\ifx\csname natexlab\endcsname\relax\def\natexlab#1{#1}\fi
\expandafter\ifx\csname bibnamefont\endcsname\relax
  \def\bibnamefont#1{#1}\fi
\expandafter\ifx\csname bibfnamefont\endcsname\relax
  \def\bibfnamefont#1{#1}\fi
\expandafter\ifx\csname citenamefont\endcsname\relax
  \def\citenamefont#1{#1}\fi
\expandafter\ifx\csname url\endcsname\relax
  \def\url#1{\texttt{#1}}\fi
\expandafter\ifx\csname urlprefix\endcsname\relax\def\urlprefix{URL }\fi
\providecommand{\bibinfo}[2]{#2}
\providecommand{\eprint}[2][]{\url{#2}}

\bibitem[{\citenamefont{O'Dell et~al.}(2000)\citenamefont{O'Dell, Giovanazzi,
  Kurizki, and Akulin}}]{ODe00}
\bibinfo{author}{\bibfnamefont{D.}~\bibnamefont{O'Dell}},
  \bibinfo{author}{\bibfnamefont{S.}~\bibnamefont{Giovanazzi}},
  \bibinfo{author}{\bibfnamefont{G.}~\bibnamefont{Kurizki}}, \bibnamefont{and}
  \bibinfo{author}{\bibfnamefont{V.~M.} \bibnamefont{Akulin}},
  \bibinfo{journal}{Phys. Rev. Lett.} \textbf{\bibinfo{volume}{84}},
  \bibinfo{pages}{5687} (\bibinfo{year}{2000}).

\bibitem[{\citenamefont{Giovanazzi
  et~al.}(2001{\natexlab{a}})\citenamefont{Giovanazzi, O'Dell, and
  Kurizki}}]{Gio01a}
\bibinfo{author}{\bibfnamefont{S.}~\bibnamefont{Giovanazzi}},
  \bibinfo{author}{\bibfnamefont{D.}~\bibnamefont{O'Dell}}, \bibnamefont{and}
  \bibinfo{author}{\bibfnamefont{G.}~\bibnamefont{Kurizki}},
  \bibinfo{journal}{Phys. Rev. A} \textbf{\bibinfo{volume}{63}},
  \bibinfo{pages}{031603(R)} (\bibinfo{year}{2001}{\natexlab{a}}).

\bibitem[{\citenamefont{Giovanazzi
  et~al.}(2001{\natexlab{b}})\citenamefont{Giovanazzi, Kurizki, Mazets, and
  Stringari}}]{Gio01b}
\bibinfo{author}{\bibfnamefont{S.}~\bibnamefont{Giovanazzi}},
  \bibinfo{author}{\bibfnamefont{G.}~\bibnamefont{Kurizki}},
  \bibinfo{author}{\bibfnamefont{I.~E.} \bibnamefont{Mazets}},
  \bibnamefont{and}
  \bibinfo{author}{\bibfnamefont{S.}~\bibnamefont{Stringari}},
  \bibinfo{journal}{Europhys. Lett.} \textbf{\bibinfo{volume}{56}},
  \bibinfo{pages}{1} (\bibinfo{year}{2001}{\natexlab{b}}).

\bibitem[{\citenamefont{Santos et~al.}(2000)\citenamefont{Santos, Shlyapnikov,
  Zoller, and Lewenstein}}]{santos00}
\bibinfo{author}{\bibfnamefont{L.}~\bibnamefont{Santos}},
  \bibinfo{author}{\bibfnamefont{G.~V.} \bibnamefont{Shlyapnikov}},
  \bibinfo{author}{\bibfnamefont{P.}~\bibnamefont{Zoller}}, \bibnamefont{and}
  \bibinfo{author}{\bibfnamefont{M.}~\bibnamefont{Lewenstein}},
  \bibinfo{journal}{Phys.\,Rev.\,Lett.} \textbf{\bibinfo{volume}{85}},
  \bibinfo{pages}{1791} (\bibinfo{year}{2000}).

\bibitem[{\citenamefont{Baranov et~al.}(2002)\citenamefont{Baranov, Dobrek,
  G{\'o}ral, Santos, and Lewenstein}}]{baranov02}
\bibinfo{author}{\bibfnamefont{M.}~\bibnamefont{Baranov}},
  \bibinfo{author}{\bibfnamefont{L.}~\bibnamefont{Dobrek}},
  \bibinfo{author}{\bibfnamefont{K.}~\bibnamefont{G{\'o}ral}},
  \bibinfo{author}{\bibfnamefont{L.}~\bibnamefont{Santos}}, \bibnamefont{and}
  \bibinfo{author}{\bibfnamefont{M.}~\bibnamefont{Lewenstein}},
  \bibinfo{journal}{Phys.\,Scr.} \textbf{\bibinfo{volume}{T102}},
  \bibinfo{pages}{74} (\bibinfo{year}{2002}).

\bibitem[{\citenamefont{G{\'o}ral et~al.}(2002)\citenamefont{G{\'o}ral, Santos,
  and Lewenstein}}]{goral02a}
\bibinfo{author}{\bibfnamefont{K.}~\bibnamefont{G{\'o}ral}},
  \bibinfo{author}{\bibfnamefont{L.}~\bibnamefont{Santos}}, \bibnamefont{and}
  \bibinfo{author}{\bibfnamefont{M.}~\bibnamefont{Lewenstein}},
  \bibinfo{journal}{Phys.\,Rev.\,Lett.} \textbf{\bibinfo{volume}{88}},
  \bibinfo{pages}{170406} (\bibinfo{year}{2002}).

\bibitem[{\citenamefont{G{\'o}ral and Santos}(2002)}]{goral02b}
\bibinfo{author}{\bibfnamefont{K.}~\bibnamefont{G{\'o}ral}} \bibnamefont{and}
  \bibinfo{author}{\bibfnamefont{L.}~\bibnamefont{Santos}},
  \bibinfo{journal}{Phys.\,Rev.\,A} \textbf{\bibinfo{volume}{66}},
  \bibinfo{pages}{023613} (\bibinfo{year}{2002}).

\bibitem[{\citenamefont{Giovanazzi et~al.}(2003)\citenamefont{Giovanazzi,
  G{\"o}rlitz, and Pfau}}]{giovanazzi03}
\bibinfo{author}{\bibfnamefont{S.}~\bibnamefont{Giovanazzi}},
  \bibinfo{author}{\bibfnamefont{A.}~\bibnamefont{G{\"o}rlitz}},
  \bibnamefont{and} \bibinfo{author}{\bibfnamefont{T.}~\bibnamefont{Pfau}},
  \bibinfo{journal}{J.\,Opt.\,B.} \textbf{\bibinfo{volume}{5}},
  \bibinfo{pages}{S208} (\bibinfo{year}{2003}).

\bibitem[{\citenamefont{Griesmaier et~al.}(2005)\citenamefont{Griesmaier,
  Werner, Hensler, Stuhler, and Pfau}}]{griesmaier05}
\bibinfo{author}{\bibfnamefont{A.}~\bibnamefont{Griesmaier}},
  \bibinfo{author}{\bibfnamefont{J.}~\bibnamefont{Werner}},
  \bibinfo{author}{\bibfnamefont{S.}~\bibnamefont{Hensler}},
  \bibinfo{author}{\bibfnamefont{J.}~\bibnamefont{Stuhler}}, \bibnamefont{and}
  \bibinfo{author}{\bibfnamefont{T.}~\bibnamefont{Pfau}},
  \bibinfo{journal}{Phys.\,Rev.\,Lett.} \textbf{\bibinfo{volume}{94}},
  \bibinfo{pages}{160401} (\bibinfo{year}{2005}).

\bibitem[{\citenamefont{Stuhler et~al.}(2005)\citenamefont{Stuhler, Griesmaier,
  Koch, Fattori, Pfau, Giovanazzi, Pedri, and Santos}}]{stuhler05}
\bibinfo{author}{\bibfnamefont{J.}~\bibnamefont{Stuhler}},
  \bibinfo{author}{\bibfnamefont{A.}~\bibnamefont{Griesmaier}},
  \bibinfo{author}{\bibfnamefont{T.}~\bibnamefont{Koch}},
  \bibinfo{author}{\bibfnamefont{M.}~\bibnamefont{Fattori}},
  \bibinfo{author}{\bibfnamefont{T.}~\bibnamefont{Pfau}},
  \bibinfo{author}{\bibfnamefont{S.}~\bibnamefont{Giovanazzi}},
  \bibinfo{author}{\bibfnamefont{P.}~\bibnamefont{Pedri}}, \bibnamefont{and}
  \bibinfo{author}{\bibfnamefont{L.}~\bibnamefont{Santos}},
  \bibinfo{journal}{Phys.\,Rev.\,Lett.} \textbf{\bibinfo{volume}{95}},
  \bibinfo{pages}{150406} (\bibinfo{year}{2005}).

\bibitem[{\citenamefont{Koch et~al.}(2008)\citenamefont{Koch, Lahaye, Metz,
  Fr{\"o}hlich, Griesmaier, and Pfau}}]{koch08}
\bibinfo{author}{\bibfnamefont{T.}~\bibnamefont{Koch}},
  \bibinfo{author}{\bibfnamefont{T.}~\bibnamefont{Lahaye}},
  \bibinfo{author}{\bibfnamefont{J.}~\bibnamefont{Metz}},
  \bibinfo{author}{\bibfnamefont{B.}~\bibnamefont{Fr{\"o}hlich}},
  \bibinfo{author}{\bibfnamefont{A.}~\bibnamefont{Griesmaier}},
  \bibnamefont{and} \bibinfo{author}{\bibfnamefont{T.}~\bibnamefont{Pfau}},
  \bibinfo{journal}{Nature Physics} \textbf{\bibinfo{volume}{4}},
  \bibinfo{pages}{218} (\bibinfo{year}{2008}).

\bibitem[{\citenamefont{Papadopoulos et~al.}(2007)\citenamefont{Papadopoulos,
  Wagner, Wunner, and Main}}]{Pap07}
\bibinfo{author}{\bibfnamefont{I.}~\bibnamefont{Papadopoulos}},
  \bibinfo{author}{\bibfnamefont{P.}~\bibnamefont{Wagner}},
  \bibinfo{author}{\bibfnamefont{G.}~\bibnamefont{Wunner}}, \bibnamefont{and}
  \bibinfo{author}{\bibfnamefont{J.}~\bibnamefont{Main}},
  \bibinfo{journal}{Phys. Rev. A} \textbf{\bibinfo{volume}{76}},
  \bibinfo{eid}{053604} (\bibinfo{year}{2007}).

\bibitem[{\citenamefont{Cartarius et~al.}(2008)\citenamefont{Cartarius, Main,
  and Wunner}}]{Car07b}
\bibinfo{author}{\bibfnamefont{H.}~\bibnamefont{Cartarius}},
  \bibinfo{author}{\bibfnamefont{J.}~\bibnamefont{Main}}, \bibnamefont{and}
  \bibinfo{author}{\bibfnamefont{G.}~\bibnamefont{Wunner}},
  \bibinfo{journal}{Phys. Rev. A} \textbf{\bibinfo{volume}{77}},
  \bibinfo{eid}{013618} (\bibinfo{year}{2008}).

\bibitem[{\citenamefont{Kato}(1966)}]{Kato66}
\bibinfo{author}{\bibfnamefont{T.}~\bibnamefont{Kato}},
  \emph{\bibinfo{title}{Perturbation theory for linear operators}}
  (\bibinfo{publisher}{Springer}, \bibinfo{address}{Berlin},
  \bibinfo{year}{1966}).

\bibitem[{\citenamefont{Heiss}(1999)}]{Hei99}
\bibinfo{author}{\bibfnamefont{W.~D.} \bibnamefont{Heiss}},
  \bibinfo{journal}{Eur. Phys. J. D} \textbf{\bibinfo{volume}{7}},
  \bibinfo{pages}{1} (\bibinfo{year}{1999}).

\bibitem[{\citenamefont{Heiss}(2004)}]{Hei04}
\bibinfo{author}{\bibfnamefont{W.~D.} \bibnamefont{Heiss}},
  \bibinfo{journal}{J. Phys. A} \textbf{\bibinfo{volume}{37}},
  \bibinfo{pages}{2455} (\bibinfo{year}{2004}).

\bibitem[{\citenamefont{G\"{u}nther et~al.}(2007)\citenamefont{G\"{u}nther,
  Rotter, and Samsonov}}]{Gue07}
\bibinfo{author}{\bibfnamefont{U.}~\bibnamefont{G\"{u}nther}},
  \bibinfo{author}{\bibfnamefont{I.}~\bibnamefont{Rotter}}, \bibnamefont{and}
  \bibinfo{author}{\bibfnamefont{B.~F.} \bibnamefont{Samsonov}},
  \bibinfo{journal}{J. Phys. A} \textbf{\bibinfo{volume}{40}},
  \bibinfo{pages}{8815} (\bibinfo{year}{2007}).

\bibitem[{\citenamefont{Cartarius et~al.}(2007)\citenamefont{Cartarius, Main,
  and Wunner}}]{Car07a}
\bibinfo{author}{\bibfnamefont{H.}~\bibnamefont{Cartarius}},
  \bibinfo{author}{\bibfnamefont{J.}~\bibnamefont{Main}}, \bibnamefont{and}
  \bibinfo{author}{\bibfnamefont{G.}~\bibnamefont{Wunner}},
  \bibinfo{journal}{Phys. Rev. Lett.} \textbf{\bibinfo{volume}{99}},
  \bibinfo{eid}{173003} (\bibinfo{year}{2007}).

\bibitem[{\citenamefont{Dirac}(1930)}]{Dir30}
\bibinfo{author}{\bibfnamefont{P.~A.~M.} \bibnamefont{Dirac}},
  \bibinfo{journal}{Proc. Cam. Phil. Soc.} \textbf{\bibinfo{volume}{26}},
  \bibinfo{pages}{376} (\bibinfo{year}{1930}).

\bibitem[{\citenamefont{McLachlan}(1964)}]{McL64}
\bibinfo{author}{\bibfnamefont{A.~D.} \bibnamefont{McLachlan}},
  \bibinfo{journal}{Mol. Phys.} \textbf{\bibinfo{volume}{8}},
  \bibinfo{pages}{39} (\bibinfo{year}{1964}).

\bibitem[{\citenamefont{P\'erez-Garc\'{\i}a
  et~al.}(1997)\citenamefont{P\'erez-Garc\'{\i}a, Michinel, Cirac, Lewenstein,
  and Zoller}}]{Per97}
\bibinfo{author}{\bibfnamefont{V.~M.} \bibnamefont{P\'erez-Garc\'{\i}a}},
  \bibinfo{author}{\bibfnamefont{H.}~\bibnamefont{Michinel}},
  \bibinfo{author}{\bibfnamefont{J.~I.} \bibnamefont{Cirac}},
  \bibinfo{author}{\bibfnamefont{M.}~\bibnamefont{Lewenstein}},
  \bibnamefont{and} \bibinfo{author}{\bibfnamefont{P.}~\bibnamefont{Zoller}},
  \bibinfo{journal}{Phys. Rev. A} \textbf{\bibinfo{volume}{56}},
  \bibinfo{pages}{1424} (\bibinfo{year}{1997}).

\bibitem[{\citenamefont{Huepe et~al.}(2003)\citenamefont{Huepe, Tuckerman,
  M\'etens, and Brachet}}]{Hue03}
\bibinfo{author}{\bibfnamefont{C.}~\bibnamefont{Huepe}},
  \bibinfo{author}{\bibfnamefont{L.~S.} \bibnamefont{Tuckerman}},
  \bibinfo{author}{\bibfnamefont{S.}~\bibnamefont{M\'etens}}, \bibnamefont{and}
  \bibinfo{author}{\bibfnamefont{M.~E.} \bibnamefont{Brachet}},
  \bibinfo{journal}{Phys. Rev. A} \textbf{\bibinfo{volume}{68}},
  \bibinfo{pages}{023609} (\bibinfo{year}{2003}).

\bibitem[{\citenamefont{Kagan et~al.}(1998)\citenamefont{Kagan, Muryshev, and
  Shlyapnikov}}]{Kag98}
\bibinfo{author}{\bibfnamefont{Y.}~\bibnamefont{Kagan}},
  \bibinfo{author}{\bibfnamefont{A.~E.} \bibnamefont{Muryshev}},
  \bibnamefont{and} \bibinfo{author}{\bibfnamefont{G.~V.}
  \bibnamefont{Shlyapnikov}}, \bibinfo{journal}{Phys. Rev. Lett.}
  \textbf{\bibinfo{volume}{81}}, \bibinfo{pages}{933} (\bibinfo{year}{1998}).

\bibitem[{\citenamefont{Donley et~al.}(2001)\citenamefont{Donley, Claussen,
  Cornish, Roberts, Cornell, and Wieman}}]{Don01}
\bibinfo{author}{\bibfnamefont{E.~A.} \bibnamefont{Donley}},
  \bibinfo{author}{\bibfnamefont{N.~R.} \bibnamefont{Claussen}},
  \bibinfo{author}{\bibfnamefont{S.~L.} \bibnamefont{Cornish}},
  \bibinfo{author}{\bibfnamefont{J.~L.} \bibnamefont{Roberts}},
  \bibinfo{author}{\bibfnamefont{E.~A.} \bibnamefont{Cornell}},
  \bibnamefont{and} \bibinfo{author}{\bibfnamefont{C.~E.}
  \bibnamefont{Wieman}}, \bibinfo{journal}{Nature}
  \textbf{\bibinfo{volume}{412}}, \bibinfo{pages}{295 } (\bibinfo{year}{2001}).

\bibitem[{\citenamefont{Broeckhove et~al.}(1989)\citenamefont{Broeckhove,
  Lathouwers, and Leuven}}]{Bro89}
\bibinfo{author}{\bibfnamefont{J.}~\bibnamefont{Broeckhove}},
  \bibinfo{author}{\bibfnamefont{L.}~\bibnamefont{Lathouwers}},
  \bibnamefont{and} \bibinfo{author}{\bibfnamefont{P.~V.}
  \bibnamefont{Leuven}}, \bibinfo{journal}{J. Phys. A}
  \textbf{\bibinfo{volume}{22}}, \bibinfo{pages}{4395} (\bibinfo{year}{1989}).

\bibitem[{\citenamefont{Feit et~al.}(1982)\citenamefont{Feit, Fleck, and
  Steiger}}]{Fei82}
\bibinfo{author}{\bibfnamefont{M.~D.} \bibnamefont{Feit}},
  \bibinfo{author}{\bibfnamefont{J.~A.} \bibnamefont{Fleck},
  \bibfnamefont{Jr.}}, \bibnamefont{and}
  \bibinfo{author}{\bibfnamefont{A.}~\bibnamefont{Steiger}},
  \bibinfo{journal}{J. Comp. Phys.} \textbf{\bibinfo{volume}{47}},
  \bibinfo{pages}{412} (\bibinfo{year}{1982}).

\bibitem[{\citenamefont{Wagner et~al.}(2008)\citenamefont{Wagner, Cartarius,
  Fab{\v c}i{\v c}, Main, and Wunner}}]{Wag08}
\bibinfo{author}{\bibfnamefont{P.}~\bibnamefont{Wagner}},
  \bibinfo{author}{\bibfnamefont{H.}~\bibnamefont{Cartarius}},
  \bibinfo{author}{\bibfnamefont{T.}~\bibnamefont{Fab{\v c}i{\v c}}},
  \bibinfo{author}{\bibfnamefont{J.}~\bibnamefont{Main}}, \bibnamefont{and}
  \bibinfo{author}{\bibfnamefont{G.}~\bibnamefont{Wunner}}
  (\bibinfo{year}{2008}), \bibinfo{note}{preprint arXiv:0802.4055}.

\end{thebibliography}

\end{document}